%% file: ht-converse-bounds.tex
\documentclass[onecolumn,draftcls]{IEEEtran}

\usepackage[latin1]{inputenc}
\usepackage[english]{babel}
\usepackage{graphicx}
\usepackage{amsmath, amsfonts, amssymb, float, latexsym, stackrel, dsfont}
\usepackage{cite}
\usepackage{balance}

\usepackage{pgfplots}
\usetikzlibrary{plotmarks}

\usepackage{notation}

\newcommand{\refE}[1]   {(\ref{eqn:#1})}
\newcommand{\refS}[1]   {Section~\ref{sec:#1}}

\begin{document}
\title{\huge Bayesian $M$-ary Hypothesis Testing:\\
       The Meta-Converse and Verd\'u-Han Bounds are Tight}
\author{Gonzalo~Vazquez-Vilar,~\IEEEmembership{Member,~IEEE,}
        Adri\`a~Tauste~Campo,~\IEEEmembership{Member,~IEEE,}\\
        Albert~Guill{\'e}n~i~F{\`a}bregas,~\IEEEmembership{Senior Member,~IEEE,}
        Alfonso~Martinez,~\IEEEmembership{Senior Member,~IEEE}%
\thanks{G. Vazquez-Vilar was with the Department of Information and Communica- tion Technologies, Universitat Pompeu Fabra, Barcelona 08018, Spain. He is now with the Signal Theory and Communications Department, Universidad Carlos III de Madrid, Legan\'es 28911, Spain (e-mail: gvazquez@ieee.org).}
\thanks{A. Tauste Campo and A. Martinez are with the Department of Information and Communication Technologies, Universitat Pompeu Fabra, Barcelona 08018, Spain (e-mail: adria.tauste@upf.edu; alfonso.martinez@ ieee.org). A. Tauste Campo is also with the Hospital del Mar Medical Research Institute, Barcelona 08003, Spain.}
\thanks{A. Guill\'en i F\`abregas is with the Instituci\'o Catalana de Recerca i Estudis Avan\c{c}ats (ICREA), with the Department of Information and Communication Technologies, Universitat Pompeu Fabra, Barcelona 08018, Spain, and also with the Department of Engineering, University of Cambridge, Cambridge CB2 1PZ, U.K. (e-mail: guillen@ieee.org).}
\thanks{This work has been funded in part by the European Research Council under ERC grant agreement 259663, by the European Union's 7th Framework Programme under grant agreements 303633 and 329837 and by the Spanish Ministry of Economy and Competitiveness under grants RYC-2011-08150, TEC2012-38800-C03-03 and FPDI-2013-18602.}
\thanks{This work was presented in part at the 2013 IEEE International Symposium on Information Theory, Istanbul, Turkey, July 7--12, 2013.}}
\maketitle
 
\begin{abstract}
Two alternative exact characterizations of the minimum error probability of Bayesian $M$-ary hypothesis testing are derived. The first expression corresponds to the error probability of an induced binary hypothesis test and implies the tightness of the meta-converse bound by Polyanskiy, Poor and Verd\'u; the second expression is function of an information-spectrum measure and implies the tightness of a generalized Verd\'u-Han lower bound. The formulas characterize the minimum error probability of several problems in information theory and help to identify the steps where existing converse bounds are loose.
\end{abstract}


\section{Introduction} 

Statistical hypothesis testing appears in areas as diverse as information theory, image processing,  signal processing, social sciences or biology. 
Depending on the field, this problem can be referred to as classification, discrimination, signal detection or model selection.
The goal of $M$-ary hypothesis testing is to decide among $M$ possible hypotheses based on the observation of a certain random variable.
In a Bayesian formulation, a prior distribution over the hypotheses is assumed, and the problem is translated into a minimization of the average error probability or its generalization, the Bayes risk.
When the number of hypotheses is $M = 2$, the problem is referred to as binary hypothesis testing. While a Bayesian approach in this case is still possible, the binary setting allows a simple formulation in terms of the two types of pairwise errors with no prior distribution over the hypotheses. The work of Neyman and Pearson~\cite{NeyPea33} established the optimum binary test in this setting. Thanks to its simplicity and robustness, this has been the most popular approach in the literature.

In the context of reliable communication, binary hypothesis testing has been instrumental in the derivation of converse bounds to the error probability. In \cite[Sec. III]{Shan67} Shannon, Gallager and Berlekamp derived lower bounds to the error probability in the transmission of $M$~messages, including the sphere-packing bound, by analyzing an instance of binary hypothesis testing~\cite{Shan67,Shan67II}. In~\cite{Forney1968}, Forney used a binary hypothesis test to determine the optimum decision regions in decoding with erasures. In~\cite{Blahut74}, Blahut emphasized the fundamental role of binary hypothesis testing in information theory and provided an alternative derivation of the sphere-packing exponent. Inspired by this result, Omura presented in~\cite{Omura1975} a general method for lower-bounding the error probability of channel coding and source coding.
More recently, Polyanskiy, Poor and Verd{\'u}~\cite{Pol09} applied the Neyman-Pearson lemma to a particular binary hypothesis test to derive the meta-converse bound, a fundamental finite-length lower bound to the channel-coding error probability from which several converse bounds can be recovered. The meta-converse bound was extended to joint source-channel coding in~\cite{allerton12,Kost13}.

The information-spectrum method expresses the error probability as the tail probability of a certain random variable, often referred to as information density, entropy density or information random variable \cite{Han03}. This idea was initially used by Shannon in \cite{Shan57} to obtain bounds to the channel coding error probability. Verd\'u and Han capitalized on this analysis to provide error bounds and capacity expressions that hold for general channels, including arbitrary memory, input and output alphabets~\cite{HanVer93,VerHan94,PoorVer95} (see also~\cite{Han03}).

In this work, we further develop the connection between hypothesis testing, information-spectrum and converse bounds in information theory by providing a number of alternative expressions for the error probability of Bayesian $M$-ary hypothesis testing. We show that this probability can be equivalently described by the error probability of a binary hypothesis test with certain parameters. In particular, this result implies that the meta-converse bound by Polyanskiy, Poor and Verd{\'u} gives the minimum error probability when it is optimized over its free parameters.
We also provide an explicit alternative expression using information-spectrum measures and illustrate the connection with existing information-spectrum bounds. This result implies that a suitably optimized generalization of the Verd\'u-Han bound also gives the minimum error probability. We discuss in some detail examples and extensions.

The rest of this paper is organized as follows. In Section~\ref{sec:BaryHT} of this paper we formalize the binary hypothesis testing problem and introduce notation.
In \refS{MaryHT} we present $M$-ary hypothesis testing  and propose a number of alternative expressions to the average error probability.
The hypothesis-testing framework is related to several previous converse results in \refS{applications}. Proofs of several results are included in the appendices.

\section{Binary Hypothesis Testing}\label{sec:BaryHT}

Let $Y$ be a random variable taking values over a discrete alphabet $\Yc$. 
We define two hypotheses $\Hc_0$ and $\Hc_1$, such that $Y$ is distributed according to a given distribution $P$ under $\Hc_0$, and according to a distribution $Q$ under $\Hc_1$.
A binary hypothesis test is a mapping $\Yc \to \{0,1\}$,  where $0$ and $1$ correspond respectively to $\Hc_0$ and $\Hc_1$.
Denoting by $\hat H \in \{0,1\}$ the random variable associated with the test output, we may describe the (possibly randomized) test by a conditional distribution $T \triangleq P_{\hat H|Y}$. 

The performance of a binary hypothesis test is characterized by two conditional error probabilities, namely $\epsilon_{0}(P,T)$ or type-0 probability, and $\epsilon_{1}(P,T)$ or type-1 probability, respectively given by
\begin{align}
 \epsilon_{0}(P,T) 
     &\triangleq \Pr\bigl[\hat H = 1 \,\big|\, \Hc_0 \bigr]
      = \sum_{y} P(y) T(1|y),
 \label{eqn:bht-type0error}\\
 \epsilon_{1}(Q,T)
     &\triangleq \Pr\bigl[\hat H = 0 \,\big|\, \Hc_1 \bigr]
      = \sum_{y} Q(y) T(0|y).
 \label{eqn:bht-type1error}
\end{align}

In the Bayesian setting, for $\Hc_i$ with prior probability $\Pr[\Hc_i]$, $i=0,1$, the smallest average error probability is
\begin{align}
  \bar\epsilon
    \triangleq \min_{T}\Bigl\{ \Pr[\Hc_0]\,\epsilon_{0}(P,T)
    + \Pr[\Hc_1]\,\epsilon_{1}(Q,T) \Bigr\}.
 \label{eqn:bht-average-error}
\end{align}
In the non-Bayesian setting, the priors $\Pr[\Hc_i]$, $i=0,1$, are unknown and the quantity $\bar\epsilon$ is not defined. Instead, one can characterize the optimal trade-off between $\epsilon_{0}(\cdot)$ and $\epsilon_{1}(\cdot)$. We define the smallest type-$0$ error $\epsilon_{0}(\cdot)$ among all tests $T$ with a type-$1$ error $\epsilon_{1}(\cdot)$ at most $\beta$ as
\begin{align}
\alpha_{\beta}\bigl(P, Q\bigr)
 \triangleq \min_{{T: \epsilon_{1}(Q, T) \leq \beta}} \Big\{ \epsilon_{0}(P, T) \Big\}.
 \label{eqn:bht-alpha}
\end{align}

The tests minimizing~\refE{bht-average-error} and~\refE{bht-alpha} have
the same form. The minimum is attained by the Neyman-Pearson test~\cite{NeyPea33},
\begin{align}\label{eqn:bht-NPtest}
  T_{\text{NP}} (0|y) =
  \begin{cases}
    1, & \text{ if }   \frac{P(y)}{Q(y)} > \gamma,\\
    p, & \text{ if }   \frac{P(y)}{Q(y)} = \gamma,\\
    0, & \text{ otherwise},
  \end{cases}
\end{align}
where $\gamma\geq 0$ and $p \in [0,1]$ are parameters.
When $\gamma = \frac{\Pr[\Hc_1]}{\Pr[\Hc_0]}$, the test $T_{\text{NP}}$
minimizes~\refE{bht-average-error} with the value of $p$
being irrelevant since it does not affect the objective.
When $\gamma$ and $p$ are chosen such that the type-$1$ error $\epsilon_{1}(Q,T_{\text{NP}})$ is equal to $\beta$, $T_{\text{NP}}$ attains the minimum in~\refE{bht-alpha}.
The test minimizing~\refE{bht-average-error} and~\refE{bht-alpha} is not unique in general, as the form of the test can vary for observations $y$ satisfying $P(y)=Q(y)$. Any test achieving~\refE{bht-alpha} is said to be optimal in the Neyman-Pearson sense.

\section{$M$-ary Hypothesis Testing}\label{sec:MaryHT}

Consider two random variables $V$ and $Y$ with joint distribution $\pvy$, where $V$ takes values on a discrete alphabet $\Vc$ of cardinality $|\Vc| = M$, and $Y$ takes values in a discrete alphabet $\Yc$. We shall assume that the cardinality $|\Vc|$ is finite; see Remark \ref{remark:mht-alpha} in \refS{proof-mht-alpha} for an extension to infinite alphabets $\Vc$. While throughout the article we use discrete notation for clarity of exposition, the results directly generalize to continuous alphabets $\Yc$; see Remark \ref{remark:proof-continuous-mht-alpha} in \refS{proof-mht-alpha}.

The estimation of $V$ given $Y$ is an $M$-ary hypothesis-testing problem. Since the joint distribution $\pvy$ defines a prior distribution $\pv$ over the alternatives, the problem is naturally cast within the Bayesian framework. 

An $M$-ary hypothesis test is defined by a (possibly random) transformation $P_{\hat V|Y} : \Yc \to \Vc$, where $\hat{V}$ denotes the random variable associated to the test output.\footnote{While both binary and $M$-ary hypothesis tests are defined by conditional distributions, to avoid confusion, we denote binary tests by $T$ and $M$-ary tests by $P_{\hat V|Y}$.} We denote the average error probability of a test $P_{\hat V|Y}$ by $\bar\epsilon(P_{\hat V|Y})$. This probability is given by 
\begin{align}
\bar\epsilon(P_{\hat V|Y})
  &\triangleq \Pr\left[\hat V \neq V\right]
  \label{eqn:mht-epsdef-1}\\
  &= 1 - \sum_{v,y} \pvy(v,y)P_{\hat V|Y}(v|y).
  \label{eqn:mht-epsdef-2}
\end{align}
Minimizing over all possible conditional distributions
$P_{\hat V|Y}$ gives the smallest average error probability, namely 
\begin{align}
  \bar\epsilon 
  &\triangleq \min_{P_{\hat V|Y}} \bar\epsilon(P_{\hat V|Y}).
  \label{eqn:mht-epsopt-1}
\end{align}
An optimum test chooses the hypothesis $v$ with largest posterior probability $P_{V|Y}(v|y)$
given the observation $y$, that is the Maximum a Posteriori (MAP) test. The MAP test that breaks ties randomly with equal probability is given by
\begin{equation}\label{eqn:mht-PMAP}
   P^{\text{MAP}}_{\hat{V}|Y}(v|y) = \begin{cases}
   \frac{1}{|\Sc(y)|}, & \text{ if }  v \in \Sc(y),\\
    0, & \text{ otherwise,}
  \end{cases}
\end{equation}
where the set $\Sc(y)$ is defined as
\begin{align} \label{eqn:mht-Sdef}
  \Sc(y) &\triangleq 
    \left\{ v \in \Vc \,\;\big|\;  P_{V|Y}(v|y) 
     = \max_{v'\in\Vc}  P_{V|Y}(v'|y) \right\}.
\end{align}
Substituting~\eqref{eqn:mht-PMAP} in~\eqref{eqn:mht-epsdef-2} gives
\begin{align}
  \bar\epsilon 
  &= 1 - \sum_{v,y} \pvy(v,y) P^{\text{MAP}}_{\hat{V}|Y}(v|y)  \label{eqn:mht-epsopt-2}\\
  &= 1- \sum_y \max_{v'} \pvy(v',y).
\label{eqn:mht-epsopt-3}
\end{align}

The next theorem introduces two alternative equivalent expressions for the minimum error probability $\bar\epsilon$.
\begin{theorem}
\label{thm:mht-alpha} 
The minimum error probability of an $M$-ary hypothesis test (with possibly non-equally likely hypotheses) can be expressed as
\begin{align} 
\bar\epsilon 
&= \max_{\qy} \alpha_{\frac{1}{M}} \bigl(\pvy, \qv \times \qy\bigr)
\label{eqn:meta}\\
&=\maxp_{\qy} \sup_{\gamma\geq 0} \left\{ \Pr\left[ \frac{\pvy(V,Y)}{ \qy(Y) } \leq \gamma \right] - \gamma \right\},
\label{eqn:tight-vh}
\end{align}
where $\qv(v) \triangleq \frac{1}{M}$ for all $v\in\Vc$, and the probability in \refE{tight-vh} is computed with respect to $\pvy$.
Moreover, a maximizing distribution $\qy$ in both expressions is 
\begin{equation} \label{eqn:qyMAP-def}
\qy^{\star}(y) \triangleq \frac{1}{\mu} \max_{v'} \pvy(v',y),
\end{equation}
 where $\mu \triangleq \sum_y \max_{v'} \pvy(v',y)$ is a normalizing constant.\end{theorem} 
\begin{IEEEproof}
See \refS{proof-mht-alpha}.
\end{IEEEproof}

Eq. \refE{meta} in Theorem~\ref{thm:mht-alpha} shows that the error probability of Bayesian $M$-ary hypothesis testing can be expressed as the best type-$0$ error probability of an induced binary hypothesis test discriminating between the original distribution $\pvy$ and an alternative product distribution $\qv \times \qy^{\star}$ with type-$1$-error equal to $\frac{1}{M}$.
Eq. \refE{tight-vh} in Theorem~\ref{thm:mht-alpha} provides an alternative characterization based on information-spectrum measures, namely the generalized information density $\log\frac{\pvy(v,y)}{\qy(y)}$. By choosing $\qy = \qy^{\star}$ and $\gamma = \mu$, the term $\Pr\left[ \frac{\pvy(V,Y)}{\qy(Y) } \leq \gamma \right] - \gamma$ can be interpreted as the error probability of an $M$-ary hypothesis test that, for each $v$, compares the posterior likelihood ${P_{V|Y}(v|y)}$ with a threshold equal to $\max_{v'} {P_{V|Y}(v'|y)}$ and decides accordingly, i.~e., this test emulates the MAP test yielding the exact error probability.
The two alternative expressions provided in Theorem~\ref{thm:mht-alpha} are not easier to compute than $\bar\epsilon$ in \refE{mht-epsopt-3}. To see this, note that the normalization factor $\mu$ in $\qy^{\star}$ is such that $\mu = 1 - \bar\epsilon$.

For any fixed test $P_{\hat V|Y}$, not necessarily MAP, using \refE{mht-epsopt-1} it follows that $\bar\epsilon(P_{\hat V|Y}) \geq \bar\epsilon$.
Therefore, Theorem~\ref{thm:mht-alpha} provides a lower bound to the error probability of any $M$-ary hypothesis test. This bound is expressed in \refE{meta} as a binary hypothesis test discriminating between $\pvy$ and an auxiliary distribution $\qvy = \qv \times \qy$. Optimizing  over general distributions $\qvy$ (not necessarily product) may yield tighter bounds for a fixed test $P_{\hat V|Y}$, as shown next.
\begin{theorem} \label{thm:mht-suboptimal}
The error probability of an $M$-ary hypothesis test $P_{\hat V|Y}$ satisfies
\begin{align} 
\bar\epsilon(P_{\hat V|Y}) &= \max_{\qvy} \alpha_{\epsilon_{1}(\qvy,P_{\hat V|Y})} \bigl(\pvy, \qvy \bigr)\label{eqn:meta-suboptimal}\\
&=\maxp_{\qvy} \sup_{\gamma\geq 0} \Biggl\{ \Pr\Biggl[ \frac{\pvy(V,Y)}{ \qvy(V,Y) } \leq \gamma \Biggr]
 - \gamma \epsilon_{1}(\qvy,P_{\hat V|Y}) \Biggr\},\label{eqn:tight-vh-suboptimal}
\end{align}
where
\begin{align} \label{eqn:eps1-def-suboptimal}
  \epsilon_{1}(\qvy,P_{\hat V|Y}) \triangleq \sum_{v,y} \qvy(v,y) P_{\hat V|Y}(v|y).
\end{align}
\end{theorem} 
\begin{IEEEproof}
Let us consider the binary test $T(0|v,y) = P_{\hat V|Y}(v|y)$. The type-$0$ and type-$1$ error probabilities of this test are $\epsilon_{0}(\pvy, T) = \bar\epsilon(P_{\hat V|Y})$ and $\epsilon_{1}(\qvy, T) = \epsilon_{1}(\qvy,P_{\hat V|Y})$ defined in \refE{eps1-def-suboptimal}, respectively. Therefore, from the definition of $\alpha_{(\cdot)}(\cdot)$ in \refE{bht-alpha} we obtain that, for any $\qvy$,
\begin{align} 
\bar\epsilon(P_{\hat V|Y}) 
   \geq \alpha_{\epsilon_{1}(\qvy,P_{\hat V|Y})} \bigl(\pvy, \qvy \bigr).
   \label{eqn:meta-suboptimal-bound}
\end{align}
For $\qvy=\pvy$, using that $\alpha_{\beta}(\pvy,\pvy) = 1 - \beta$, the right-hand side of \refE{meta-suboptimal-bound} becomes $1 - \epsilon_{1}(\pvy,P_{\hat V|Y})$. As $1 - \epsilon_{1}(\pvy,P_{\hat V|Y}) = 1 - \epsilon_{1}(\pvy,T) = \epsilon_{0}(\pvy,T) = \bar\epsilon(P_{\hat V|Y})$, then \refE{meta-suboptimal} follows from optimizing \refE{meta-suboptimal-bound} over $\qvy$.
To obtain~\refE{tight-vh-suboptimal} we apply the lower bound in Lemma~\ref{lem:alpha-relax-2} in \refS{proof-mht-alpha} to \refE{meta-suboptimal} and note that, for $\gamma=1$, $\qvy=\pvy$, the bound holds with equality.
\end{IEEEproof}

The proof of Theorem~\ref{thm:mht-suboptimal} shows that the auxiliary distribution $\qvy=\pvy$ maximizes \refE{meta-suboptimal} and \refE{tight-vh-suboptimal}  for any $M$-ary hypothesis test $P_{\hat V|Y}$. Nevertheless, the auxiliary distribution optimizing \refE{meta-suboptimal} and \refE{tight-vh-suboptimal} is is not unique in general, as seen in Theorem~\ref{thm:mht-alpha} for the MAP test and in the next result for arbitrary maximum-metric tests.

Consider the maximum-metric test $P^{(q)}_{\hat{V}|Y}$ that chooses the hypothesis $v$ with largest metric $q(v,y)$, where $q(v,y)$ is an arbitrary function of $v$ and $y$. This test can be equivalently described as
\begin{equation}\label{eqn:mht-PMAP-MM}
   P^{(q)}_{\hat{V}|Y}(v|y) = \begin{cases}
   \frac{1}{\left|\Sc_q(y)\right|}, & \text{ if }  v \in \Sc_q(y),\\
    0, & \text{ otherwise,}
  \end{cases}
\end{equation}
where the set $\Sc_q(y)$ is defined as
\begin{align} \label{eqn:mht-Sdef-MM}
  \Sc_q(y) &\triangleq 
    \left\{ v \in \Vc \;\Big|\;  q(v,y) = \max_{v'\in \Vc}  q(v',y) \right\}.
\end{align}

\begin{corollary} \label{cor:mht-mm}
For the maximum metric test $P_{\hat V|Y} = P_{\hat V|Y}^{(q)}$,
a distribution $\qvy$ maximizing \refE{meta-suboptimal} and \refE{tight-vh-suboptimal}
is
\begin{align}\label{eqn:qvy-def-mm}
 \qvy^{(q)}(v,y) \triangleq \frac{\pvy(v,y)}{\mu'} \frac{\max_{v'} q(v',y)}{q(v,y)},
\end{align}
where $\mu'$ is a normalizing constant.
\end{corollary} 
\begin{IEEEproof}
See Appendix \ref{apx:mht-mm}.
\end{IEEEproof}

The expressions in Theorem~\ref{thm:mht-suboptimal} still depend on the specific test through $\epsilon_{1}(\cdot)$, cf. \refE{eps1-def-suboptimal}. For the optimal MAP test, i.~e., a maximum metric test with metric $q(v,y) = P_{V|Y}(v|y)$, we obtain $\qvy^{(q)} = \qv \times \qy^{\star}$ with uniform $\qv$ and $\qy^{\star}$ defined in~\refE{qyMAP-def}. For uniform $\qv$ it holds that
\begin{align}
\epsilon_{1}(\qv \times \qy, P_{\hat V|Y}) = \frac{1}{M},
\end{align}
for any $\qy$, $P_{\hat V|Y}$. As a result, for the optimal MAP test, the expressions in Theorem~\ref{thm:mht-suboptimal} and the distribution defined in Corollary~\ref{cor:mht-mm} recover those in Theorem~\ref{thm:mht-alpha}.

\subsection{Example}\label{sec:example}
To show the computation of the various expressions in Theorem~\ref{thm:mht-alpha} let us consider the ternary hypothesis test examined in \cite[Figs. 1 and 2]{PoorVer95} and revisited in \cite[Sec. III.A]{ChenAla2012}. Let $\Vc = \Yc = \{0,1,2\}$, $\pv(v)=\frac{1}{3}$, $v=0,1,2$, and 
\begin{align}  \label{eqn:ex1-pyx}
\pyv(y|v) =\begin{cases} 0.40,& (v,y) = (0,0), (1,1) \text{ and } (2,2),\\
             0.33,& (v,y) = (0,2), (1,2) \text{ and } (2,0), \\
             0.27,& \text{otherwise}.\end{cases}
\end{align}
Direct calculation shows that the MAP estimate is $\hat{v}(y) = y$, and from \refE{mht-epsopt-3} we obtain $\bar{\epsilon}=0.6$.

In order to evaluate the expressions in Theorem~\ref{thm:mht-alpha} we first compute $\qy^{\star}$ in \refE{qyMAP-def}, which yields $\qy^{\star}(y) = \frac{1}{3}$, $y=0,1,2$. According to \refE{meta} a binary hypothesis test between $\pvy$ and $\qvy^{\star}$, where $\qvy^{\star}(v,y)=\frac{1}{9}$, for all $v,y$, with type-$1$ error $\epsilon_1=\frac{1}{3}$, yields the minimum error probability
\begin{align}
  \bar\epsilon &= \alpha_{\frac{1}{3}} \bigl(\pvy, \qvy^{\star}\bigr).
  \label{eqn:ex1-alpha}
\end{align}
Solving the Neyman-Pearson test in \refE{bht-NPtest} for the type-$1$ error $\epsilon_1=\frac{1}{3}$, we obtain $\gamma = 1.2$ and $p = 1$ and therefore
\begin{align}
\label{eqn:ex1-TNP}
  T_{\text{NP}} (0|y) =
  \begin{cases}
    1, & \text{ if }   \pvy(v,y) \geq \frac{2}{15},\\
    0, & \text{ otherwise}.
  \end{cases}
\end{align}
Hence, \refE{ex1-alpha} yields
\begin{align}
  \bar\epsilon &= \epsilon_{0}(\pvy,T_{\text{NP}}) \\
               &= 1 - \sum_{v,y} \pvy(v,y) T_{\text{NP}}(0|y) = 0.6.
\end{align}

Similarly, to evaluate \refE{tight-vh} in Theorem~\ref{thm:mht-alpha}, 
we substitute $\qy^{\star}$ to obtain
\begin{align} \label{eqn:ex1-tight-vh}
\bar\epsilon 
 &=\sup_{\gamma\geq 0} \left\{ \Pr\left[ \pvy(V,Y) \leq \frac{\gamma}{3} \right]
   - \gamma \right\}.
\end{align}

\begin{figure}[t]
    \centering
	\input{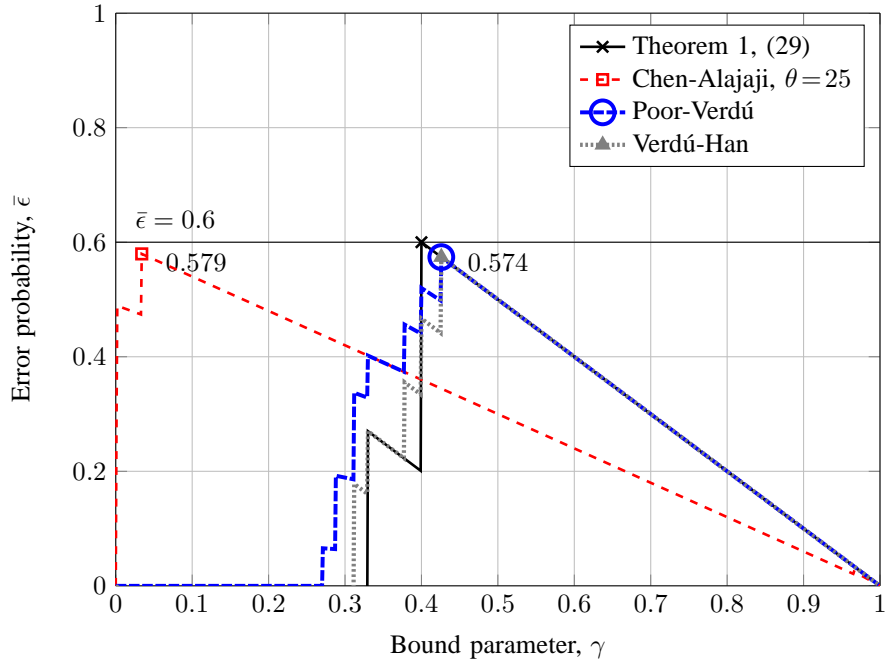}
	\caption{Information-spectrum lower bounds to the minimum error probability for the example in \refS{example}, as a function of the bound parameter~$\gamma$.}	\label{fig:ex1-bounds}
\end{figure}

Fig. \ref{fig:ex1-bounds} shows the argument of \refE{ex1-tight-vh} with respect to $\gamma \in [0,1]$ compared to the exact error probability $\bar\epsilon$, shown in the plot with an horizontal line. For comparison, we also include the Verd\'u-Han lower bound~\cite[Th.~4]{VerHan94}, 
the Poor-Verd\'u lower bound~\cite[Th.~1]{PoorVer95} 
and the lower bound proposed by Chen and Alajaji in~\cite[Th.~1]{ChenAla2012}. The Chen-Alajaji bound~\cite[Th.~1]{ChenAla2012} is parametrized by $\theta \geq 0$ and, for $\theta=1$, it reduces to the Poor-Verd\'u lower bound. We observe that \refE{ex1-tight-vh} gives the exact error probability $\bar\epsilon = 0.6$ at $\gamma=1-\bar\epsilon$. The Verd\'u-Han and the Poor-Verd\'u lower bounds both coincide and yield $\bar\epsilon \geq 0.574$. For this example, as shown in~\cite{ChenAla2012}, the Chen-Alajaji lower bound is tight for $\theta \to \infty$. For $\theta = 25$ the bound is still $\bar\epsilon \geq 0.579$.

As an application of Theorem~\ref{thm:mht-suboptimal} and Corollary~\ref{cor:mht-mm} we study now a variation of the previous example. For a hypothesis $v \in \Vc$, let $(y_1,y_2) \in \Yc^2$ denote two independent observations of the random variable $Y$ distributed according to $P_{Y|V=v}$ in \refE{ex1-pyx}. We consider the suboptimal hypothesis test that decides on the source message $v$ maximizing the metric $q(v,y_1,y_2) = \pyv(y_1|v)$. That is, for equiprobable hypotheses, this test applies the MAP rule based on the first observation, ignoring the second one.
The expressions in Theorem~\ref{thm:mht-alpha} do not depend on the decoder and yield the MAP error probability $\bar{\epsilon}=0.592$.
Then, for $P^{(q)}_{\hat{V}|Y_1Y_2}$ in \refE{mht-PMAP-MM}, it holds that $\bar\epsilon\bigl(P^{(q)}_{\hat{V}|Y_1Y_2}\bigr) \geq 0.592$.

Let us choose the auxiliary distribution
\begin{align}\label{eqn:Qex2}
  Q_{VY_1Y_2}(v, y_1, y_2) = \frac{1}{9} \pyv(y_2|v).
\end{align}
Using that $P^{(q)}_{\hat{V}|Y_1Y_2}(v|y_1,y_2) = \openone\bigl\{v = y_1\bigr\}$ is independent of~$y_2$, we obtain
\begin{align} 
\epsilon_{1}\bigl(Q_{VY_1Y_2},P^{(q)}_{\hat{V}|Y_1Y_2}\bigr)
  &= \frac{1}{9} \sum_{v,y_1,y_2} \pyv(y_2|v)  P^{(q)}_{\hat{V}|Y_1Y_2}(v | y_1,y_2)\\
  &= \frac{1}{9} \sum_{v,y_1}  \openone\bigl\{v = y_1\bigr\}\\
  &= \frac{1}{3}.
\end{align}
Therefore, the bound implied in Theorem~\ref{thm:mht-suboptimal} for this specific choice of $Q_{VY_1Y_2}$ yields
\begin{align}\label{eqn:ex2-alpha}
  \bar\epsilon\Bigl(P^{(q)}_{\hat{V}|Y_1Y_2}\Bigr)
    &\geq \alpha_{\frac{1}{3}} \bigl(P_{VY_1Y_2}, Q_{VY_1Y_2}\bigr).
\end{align}
Since the marginal corresponding to $Y_2$ is the same for $P_{VY_1Y_2}$ and $Q_{VY_1Y_2}$ in \refE{Qex2}, this component does not affect to the binary test and can be eliminated from \refE{ex2-alpha}. Therefore, the right-hand side in \refE{ex2-alpha}  coincides with that of \refE{ex1-alpha}, and yields the lower bound $\bar\epsilon\bigl(P^{(q)}_{\hat{V}|Y_1Y_2}\bigr) \geq 0.6$.
It can be checked that an application of \refE{tight-vh-suboptimal} in Theorem~\ref{thm:mht-suboptimal} yields the same result. We conclude that allowing joint distributions $Q_{VY_1Y_2}$ we obtain decoder-specific bounds. 

\subsection{Proof of Theorem~\ref{thm:mht-alpha}}\label{sec:proof-mht-alpha}
We first prove the equality between the left- and right-hand sides of \refE{meta} by showing the equivalence of the optimization problems \refE{mht-epsopt-1} and \refE{meta}.
From \refE{mht-epsopt-1} we have that
\begin{align}
  \bar\epsilon
  &= \min_{ P_{\hat V|Y}:\sum_{v} P_{\hat V|Y}(v|y) \leq 1,  y\in\Yc }  \sum_{v,y}\!\pvy(v,y)\!\left(1\!-\!P_{\hat V|Y}(v|y)\right)
  \label{eqn:mht-epsopt-4}\\
  &= \max_{\lambda(\cdot) \geq 0} \min_{ P_{\hat V|Y} } \Biggl\{ \sum_{v,y} \pvy(v,y) \left(1-P_{\hat V|Y}(v|y)\right)
   + \sum_y  \lambda(y) \left(  \sum_{v} P_{\hat V|Y}(v|y) - 1 \right) \Biggr\},
  \label{eqn:mht-epsopt-5}
\end{align}
where in \refE{mht-epsopt-4} we wrote explicitly the (active) constraints resulting from $P_{\hat V|Y}$ being a conditional distribution; and \refE{mht-epsopt-5} follows from introducing the constraints into the objective via the Lagrange multipliers $\lambda(y) \geq 0$, $y\in\Yc$.

Similarly, we write \refE{meta} as
\begin{align}
\max_{\qy} \; &\alpha_{\frac{1}{M}}\left(\pvy, \qv \times \qy\right)\notag\\
  &= \max_{\qy} \,\min_{T: \sum_{v,y} \frac{1}{M} \qy(y) T(0|v,y) \leq \frac{1}{M}} \Biggl\{\sum_{v,y} \pvy(v,y)
     T(1|v,y) \Biggr\}
  \label{eqn:mht-alphaopt-6}\\
  &= \max_{\eta\geq 0}\max_{\qy} \min_{ T} \Biggl\{\sum_{v,y} \pvy(v,y) \Bigl(1-T(0|v,y)\Bigr)
  + \eta \left( \sum_{v,y} \qy(y) T(0|v,y) - 1 \right)\Biggr\},
  \label{eqn:mht-alphaopt-7}
\end{align}
where in \refE{mht-alphaopt-6} we used the definitions of $\qv$ and $\alpha_{\beta}(\cdot)$; and \refE{mht-alphaopt-7} follows from introducing the constraint into the objective via the Lagrange multiplier $\eta$.

Since $\eta$ and $\qy$ only appear in the objective function of \refE{mht-alphaopt-7} as $\eta \qy(y)$, $y\in\Yc$, we may optimize \refE{mht-alphaopt-7} over $\bar\lambda(y) \triangleq \eta\qy(y)$ instead. Then, \refE{mht-alphaopt-7} becomes
\begin{align}
\max_{\bar\lambda(\cdot) \geq 0} \min_{T} \Biggl\{\sum_{v,y} \pvy(v,y) \Bigl(1-T(0|v,y)\Bigr)
   + \sum_{y} \bar\lambda(y) \left( \sum_{v} T(0|v,y) - 1 \right)\Biggr\}.
  \label{eqn:mht-alphaopt-8}
\end{align}

Comparing \refE{mht-epsopt-5} and \refE{mht-alphaopt-8}, it is readily seen that the optimization problems \refE{mht-epsopt-1} and \refE{meta} are equivalent. Hence, the first part of the theorem follows.

We need the following result to prove identity \eqref{eqn:tight-vh}.
\begin{lemma}\label{lem:alpha-relax-2}
For any pair of distributions $\{P,Q\}$ over $\Yc$ and any $\gamma' \geq 0$, it holds
\begin{align}
\alpha_{\beta}\bigl(P, Q\bigr)
  \geq \PP\left[\frac{P(Y)}{Q(Y)} \leq \gamma'\right]-\gamma'\beta.
\label{eqn:alpha-relax-1}
\end{align}
\end{lemma}
\begin{IEEEproof}
The bound \refE{alpha-relax-1} with the term $\PP\Bigl[\frac{P(Y)}{Q(Y)} \leq \gamma'\Bigr]$ replaced by $\PP\Bigl[\frac{P(Y)}{Q(Y)} < \gamma'\Bigr]$ corresponds to \cite[Eq. (102)]{Pol09}. The proof of the lemma follows the steps in \cite[Eq. (2.71)-(2.74)]{PolThesis} and is included in Appendix \ref{apx:alpha-relax} for completeness.
\end{IEEEproof}

Applying \refE{alpha-relax-1} to \refE{meta} with $\gamma' = \gamma M$, \mbox{$P \leftarrow \pvy$} and \mbox{$Q \leftarrow \qv \times \qy$} and optimizing over $\gamma$ we obtain
\begin{align}
  \bar\epsilon \geq \maxp_{\qy}   \sup_{\gamma\geq 0} \left\{ \Pr\left[ \frac{\pvy(V,Y)}{\qy(Y) } \leq \gamma \right] - \gamma \right\}.
\label{eqn:infspectrum}
\end{align}
By using the distribution $\qy = \qy^{\star}$ in \refE{qyMAP-def} and by choosing $\gamma = \mu$, the probability term in \refE{infspectrum} becomes
\begin{align}
\Pr\left[ \frac{\pvy(V,Y)}{ \qy^{\star}(Y) } \leq \mu \right]
=
\Pr\left[ P_{V|Y}(V|Y) \leq \max_{v'} P_{V|Y}(v'|Y) \right] = 1. \label{eqn:tight-infspectrum}
\end{align}
Substituting $\qy = \qy^{\star}$, $\gamma = \mu$, and using \refE{tight-infspectrum} in \refE{infspectrum} we obtain
\begin{align}
  \bar\epsilon
    &\geq \maxp_{\qy}   \sup_{\gamma\geq 0} \left\{ \Pr\left[ \frac{\pvy(V,Y)}{\qy(Y)} \leq \gamma \right] - \gamma \right\} \label{eqn:tight-infspectrum-1}\\
    &\geq 1-\mu \label{eqn:tight-infspectrum-2}\\
    &= 1 - \sum_{y} \max_{v'} \pvy(v',y) \label{eqn:tight-infspectrum-3}\\
    &=\bar\epsilon, \label{eqn:tight-infspectrum-4}
\end{align}
where in \refE{tight-infspectrum-3} we used the definition of $\mu$  and \refE{tight-infspectrum-4} follows from \refE{mht-epsopt-3}. 
The identity \eqref{eqn:tight-vh} in the theorem is due to \refE{tight-infspectrum-1}-\refE{tight-infspectrum-4}, where it is readily seen that $\qy = \qy^{\star}$ is a maximizer of \eqref{eqn:tight-vh}. Moreover, since $\qy^{\star}$ is a maximizer of \eqref{eqn:tight-vh}, and Lemma~\ref{lem:alpha-relax-2} applies for a fixed $\qy$, it follows that $\qy^{\star}$ is also an optimal solution to \refE{meta}. The second part of the theorem thus follows from \refE{tight-infspectrum-1}-\refE{tight-infspectrum-4}.

\begin{remark}\label{remark:mht-alpha}A simple modification of Theorem~\ref{thm:mht-alpha} generalizes the result to countably infinite alphabets $\Vc$. We define $\barqv$ to be the counting measure, i. e., $\barqv(v) = 1$ for all $v$.
The function $\alpha_{\beta}(\cdot)$ in \refE{bht-alpha} is defined for arbitrary $\sigma$-finite measures, not necessarily probabilities. Then, by substituting
$\qv$ by $\barqv$, the type-$1$ error measure is $\epsilon_{1}(\barqv \times \qy, T)=1$ for any $T$, and \refE{meta} becomes
\begin{equation}
\bar\epsilon = \max_{\qy} \alpha_{1}
            \left(\pvy, \barqv \times \qy\right).
\label{eqn:meta-bis}
\end{equation}
Since \refE{tight-vh} directly applies to both  finite or countably infinite $\Vc$, so does Theorem~\ref{thm:mht-alpha} with \refE{meta} replaced by~\refE{meta-bis}.
\end{remark}

\begin{remark}\label{remark:proof-continuous-mht-alpha}
For continuous observation alphabets $\Yc$, the constraint of $P_{\hat V|Y}$ being a
conditional distribution 
\begin{align}
  \sum_{v} P_{\hat V|Y}(v|y) \leq 1, \ y \in \Yc,
  \label{eqn:mht-const-v1}
\end{align}
can be equivalently described as 
\begin{align}
 \max_{\qy} \int \sum_{v} P_{\hat{V}|Y}(v|y) \diff \qy(y) \leq 1.
  \label{eqn:mht-const-v2}
\end{align}
The fact that \refE{mht-const-v1} implies \refE{mht-const-v2} trivially follows by averaging both sides of \refE{mht-const-v1} over an arbitrary $\qy$, and in particular, for the one maximizing \refE{mht-const-v2}.
To prove that \refE{mht-const-v2} implies \refE{mht-const-v1}, let us assume that \refE{mht-const-v1} does not hold, i.~e., $\sum_{v} P_{\hat V|Y}(v|\bar y) > 1$ for some $\bar y\in\Yc$.  Let $\bar Q_Y$ be the distribution that concentrates all the mass at $\bar y$. Since for $\qy=\bar Q_Y$ the condition \refE{mht-const-v2} is violated, so happens for the maximizing $\qy$. As a result, \refE{mht-const-v2} implies \refE{mht-const-v1}, as desired, and the equivalence between both expressions follows.

By using \refE{mht-const-v2} instead of \refE{mht-const-v1} in \refE{mht-epsopt-4}-\refE{mht-epsopt-5}, and after replacing the sums by integrals where needed, we obtain
\begin{align}
  \bar\epsilon 
  &= \max_{\eta \geq 0} \min_{ P_{\hat V|Y} } \Biggl\{ \int \sum_{v} P_{V|Y}(v|y) \left(1-P_{\hat V|Y}(v|y)\right) \diff \py(y)
  +  \eta \left( \max_{\qy} \int \sum_{v} P_{\hat{V}|Y}(v|y) \diff \qy(y) - 1\right) \Biggr\}. \label{eqn:mht-epsopt-5-bis}
\end{align}
For fixed $\qy$ the argument in \refE{mht-epsopt-5-bis} is linear with respect to $P_{\hat V|Y}$, and for fixed $P_{\hat V|Y}$ is linear with respect to $\qy$. Therefore, applying Sion's minimax theorem \cite[Cor. 3.5]{Sion58} to interchange $\min_{ P_{\hat V|Y} }$ and $\max_{\qy}$, \refE{mht-epsopt-5-bis} becomes \refE{mht-alphaopt-7}. The first part of the theorem thus holds for continuous alphabets $\Yc$.
Since Lemma \ref{lem:alpha-relax-2} applies to arbitrary probability spaces, so does \refE{infspectrum}. Therefore, for continuous alphabets $\Yc$, the second part of the theorem follows from \refE{infspectrum}, \refE{tight-infspectrum} and \refE{tight-infspectrum-1}-\refE{tight-infspectrum-4} after replacing the sum by an integral in \refE{tight-infspectrum-3}.
\end{remark}

\begin{remark}
The optimality of $\qy^{\star}$ in \refE{meta} can also be proved constructively. Consider the binary hypothesis testing problem between $\pvy$ and $\qv \times \qy^{\star}$. We define a test 
\begin{align}\label{eqn:mht-TMAP-proof}
  T_{\text{MAP}}(0|v,y) \triangleq 
  \begin{cases}
    \frac{1}{|\Sc(y)|}, & \text{ if } v \in \Sc(y),\\
    0, & \text{ otherwise.}
  \end{cases}
\end{align}
For $\qv$ uniform, the type-$1$ error probability of this test is $\epsilon_{1}(\qv \times \qy^{\star}, T_{\text{MAP}}) = \frac{1}{M}$.
Using that the MAP test is a maximum metric test with $q(v,y) = \pvy(v,y)$, according to the proof of Corollary \ref{cor:mht-mm} in Appendix~\ref{apx:mht-mm}, the type-$0$ error probability of $T_{\text{MAP}}$ is precisely  $\alpha_{\frac{1}{M}} \bigl(\pvy, \qv \times \qy^{\star}\bigr)$. 
Moreover, since $\bar\epsilon = \epsilon_{0}(\pvy, T_{\text{MAP}})$ we conclude that $\qy = \qy^{\star}$ is an optimizer of \refE{meta}. While both $T_{\text{MAP}}$ and $T_{\text{NP}}$ attain the Neyman-Pearson performance, in general they are not the same test, as they may differ in the set of points that lead to a MAP test tie, i.e., the values of $y$ such that $|\Sc(y)|>1$.
\end{remark}

\section{Connection to Previous Converse Results}\label{sec:applications}

We next study the connection between Theorem~\ref{thm:mht-alpha} and previous converse results in the literature:

\subsubsection{The meta-converse bound}
In channel coding,  one of $M$ equiprobable messages is to be sent over a  channel with one-shot law $\pyx$. The encoder maps the source message $v\in\{1,\ldots,M\}$ to a codeword $x(v)$ using a specific codebook~$\Cc$. Since there is a codeword for each message, the distribution $\pv$ induces a distribution $\px^{\Cc}$ over the channel input. At the decoder, the decision among the $M$ possible transmitted codewords based on the channel output $y$ is equivalent to an $M$-ary hypothesis test with equiprobable hypotheses. The smallest error probability of this test for a codebook $\Cc$ is denoted as $\bar\epsilon(\Cc)$.

Fixing an arbitrary $\qy$ in \eqref{eqn:meta} and considering the codeword set instead of the message set, we obtain
\begin{align}\label{eqn:mht-alpha-bound-2}
  \bar\epsilon(\Cc) 
    \geq \alpha_{\frac{1}{M}}
            \bigl(\px^{\Cc}\times\pyx, \px^{\Cc} \times \qy\bigr),
\end{align}
namely the meta-converse bound of~\cite[Th.~26]{Pol09} for a given codebook and the choice $\qxy = \px^{\Cc} \times \qy$. Theorem \ref{thm:mht-alpha} thus shows that the meta-converse bound is tight for a fixed codebook after optimization over the auxiliary distribution~$\qy$. 

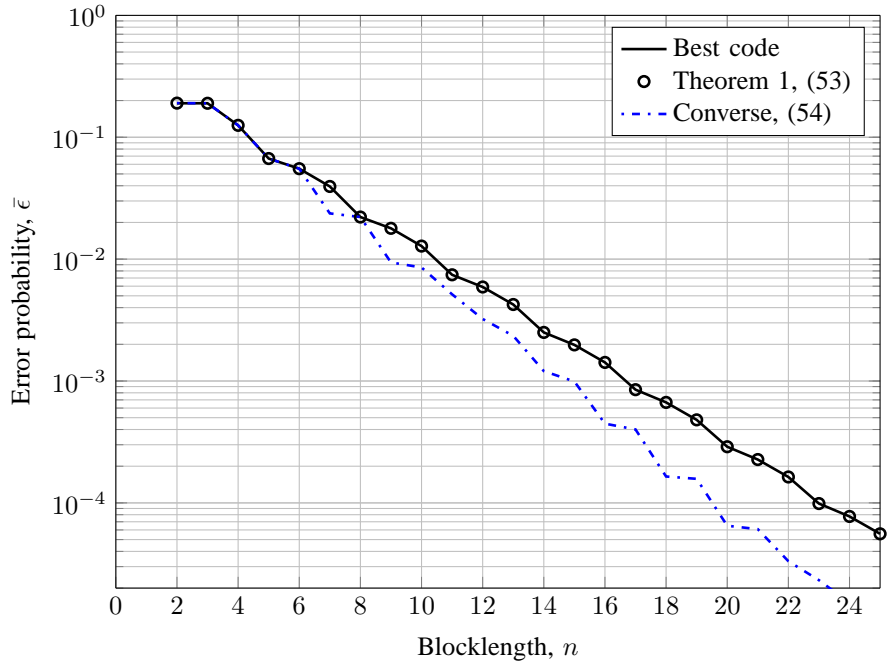
\begin{figure}[t]
	\centering
	\input{figs/bounds-cc.tikz}
	\caption{Channel coding error probability bounds for a BSC with cross-over probability  $0.1$ and $M=4$ codewords.}\label{fig:BSC-CC-bounds}
\end{figure}

Upon optimization over $\qy$ and minimization over codebooks we obtain
\begin{align}
   \min_{\Cc} \bar\epsilon(\Cc)
         &= \min_{\px^{\Cc}} \max_{\qy} \left\{
                 \alpha_{\frac{1}{M}} \bigl(\px^{\Cc}\!\times\!\pyx,
                   \px^{\Cc}\!\times\!\qy \bigr) \right\}
            \label{eqn:cc-bound-1}\\
              &\geq \min_{\px} \max_{\qy} \left\{
                 \alpha_{\frac{1}{M}} \bigl(\px\!\times\!\pyx, \px\!\times\!\qy \bigr)\right\}.
            \label{eqn:cc-bound-2}
\end{align}
The minimization in \refE{cc-bound-1} is done over the set of distributions induced by all possible codes, while the minimization in \refE{cc-bound-2} is done over the larger set of all possible distributions over the channel inputs. The bound in \refE{cc-bound-2} coincides with \cite[Th.~27]{Pol09}.

Fig. \ref{fig:BSC-CC-bounds} depicts the minimum error probability for the transmission of $M=4$ messages over $n$ independent, identically distributed channel uses of a memoryless binary symmetric channel (BSC) with single-letter cross-over probability $0.1$. We also include the meta-converse \refE{cc-bound-1}, computed for the best code~\cite[Th.~37]{moser2013} and $\qy=\qy^{\star}$, and the lower bound in \refE{cc-bound-2}. Here, we exploited the fact that for the BSC the  saddlepoint in \refE{cc-bound-2} is attained for uniform $\px, \qy$ \cite[Th.~22]{Pol13}. The computation of  \refE{cc-bound-1} and \refE{cc-bound-2} follows similar steps to those presented  in \refS{example} for a different example.
It is interesting to observe that while \refE{cc-bound-1} characterizes the exact error probability, the weakening \refE{cc-bound-2} yields a much looser bound.

\subsubsection{Lower bound based on a bank of $M$ binary tests}
Eq. \eqref{eqn:meta} relates the error probability $\bar\epsilon$ to the type-$0$ error probability of a binary test between distributions $\pvy$ and $\qv^\star\times\qy$. Instead of a single binary test, it is also possible to consider a bank of $M$ binary hypothesis tests between distributions $P_{Y|V=v}$ and $\qy$~\cite{allerton12}. In this case, we can also express the average error probability of $M$-ary hypothesis testing as
\begin{equation}
\bar\epsilon=\max_{\qy} \left\{ \sum_v \pv(v) \,\alpha_{Q_{\hat{V}}^{\star}(v)} \bigl(P_{Y|V=v}, \qy\bigr) \right\}
\label{eqn:multi}
\end{equation}
where $Q_{\hat{V}}^{\star}(v) \triangleq \sum_{y}\qy(y) P_{\hat V|Y}^{\text{MAP}}(v|y)$;
see Appendix \ref{apx:Marymultiple}.

If instead of fixing $Q_{\hat{V}}^{\star}$, we minimize \refE{multi} with respect to an arbitrary $Q_{\hat{V}}$, \refE{multi} then recovers the converse bound~\cite[Lem.~2]{allerton12} for almost-lossless joint source-channel coding.
This lower bound is not tight in general as the minimizing distribution $Q_{\hat{V}}$ need not coincide with the distribution induced by the MAP decoder.

\subsubsection{Verd\'u-Han lower bound}
Weakening the identity in \eqref{eqn:tight-vh} for an arbitrary $\qy$ we obtain
\begin{align}
  \bar\epsilon \geq \sup_{\gamma\geq 0} \left\{ \Pr\left[ \frac{ \pvy(V,Y)}{  \qy(Y) } \leq \gamma \right] - \gamma \right\}.
\label{eqn:vh}
\end{align}
By choosing $\qy=\py$ in \refE{vh} we recover the Verd\'u-Han lower bound in the channel~\cite[Th.~4]{VerHan94} and joint source-channel coding settings~\cite[Lem.~3.2]{Han07Joint}. The bound \eqref{eqn:vh} with arbitrary $\qy$ coincides with the Hayashi-Nagaoka lemma for classical-quantum channels~\cite[Lem.~4]{hayashi2003}, with its proof steps following exactly those of \cite[Th.~4]{VerHan94}. Theorem~\ref{thm:mht-alpha} shows that, by properly choosing $\qy$, this bound is tight in the classical setting.

\subsubsection{Wolfowitz's strong converse}
If we consider the hypothesis $v$ with smallest error probability in \eqref{eqn:tight-vh}, i.~e.,
\begin{align}
  \bar\epsilon
  &=
    \maxp_{\qy} \sup_{\gamma\geq 0} \left\{ \sum_{v} \pv(v) \Pr\!\left[ \frac{ \pyv(Y|v) \pv(v)}{  \qy(Y) }\!\leq\!\gamma \right]- \gamma \right\}    \label{eqn:mht-wolfowitz-pre}\\
  &\geq
    \maxp_{\qy} \sup_{\gamma\geq 0} \inf_v \left\{ \Pr\left[ \frac{ \pyv(Y|v) \pv(v)}{  \qy(Y) } \leq \gamma \right] - \gamma \right\},
    \label{eqn:mht-wolfowitz}
\end{align}
we recover Wolfowitz's channel coding strong converse~\cite{Wolf68}.
Hence, this converse bound is tight as long as the bracketed term in \refE{mht-wolfowitz} does not depend on $v$ for the pair $\{\qy, \gamma\}$ optimizing \refE{mht-wolfowitz-pre}.

\subsubsection{Poor-Verd\'u lower bound}
By applying the following lemma, we recover the Poor-Verd\'u lower bound~\cite{PoorVer95} from Theorem~\ref{thm:mht-alpha}.
Let us denote by $\PP[\Ec]$ (resp. $\QQ[\Ec]$) the probability of the event $\Ec$ with respect to the underlying distribution $P$ (resp. $Q$).
\begin{lemma}\label{lem:alpha-relax-1}
For a pair of discrete distributions $\{P,Q\}$ defined over $\Yc$ and any $\gamma' \geq 0$, such that
\begin{align}
  0\leq\beta\leq 
\frac{\QQ\left[\frac{P(Y)}{Q(Y)}>\gamma'\right]}{ \PP\left[\frac{P(Y)}{Q(Y)}>\gamma'\right]},
\label{eqn:alpha-relax-assump}
\end{align}
the following result holds,
\begin{align}
\alpha_{\beta}\bigl(P, Q\bigr)
  \geq (1-\gamma'\beta) \PP\left[\frac{P(Y)}{Q(Y)} \leq \gamma'\right].
\label{eqn:alpha-relax-2}
\end{align}
\end{lemma}
\begin{IEEEproof}
See Appendix \ref{apx:alpha-relax}.
\end{IEEEproof}

Using Lemma~\ref{lem:alpha-relax-1} with $\gamma' = \gamma M$, $P \leftarrow \pvy$ and $Q \leftarrow \qv \times \qy$ where $\qv$ is uniform, via \refE{meta}, we obtain
\begin{align}
  \bar\epsilon \geq
    (1-\gamma) \Pr\left[ \frac{ \pvy(V,Y)}{ \qy(Y) } \leq \gamma \right],
\label{eqn:pv}
\end{align}
provided that $\qy$ and $\gamma\geq0$ satisfy
\begin{align}
  {\sum_{v,y} \pvy(v,y)
            \openone\left\{\frac{ \pvy(v,y)}{ \qy(y)}>\gamma\right\}}
       \leq
       {\sum_{v,y} \qy(y)
            \openone\left\{\frac{ \pvy(v,y)}{ \qy(y)}>\gamma\right\}}.\label{eqn:pv_cond}
\end{align}
This condition is fulfilled for any $\gamma \geq 0$ if $\qy=\py$ or $\qy = \qy^{\star}$ as defined in \refE{qyMAP-def}.
However, there exist pairs $\{\gamma,\qy\}$ for which \refE{pv_cond} does not hold.
For $\qy=\py$, and optimizing over $\gamma\geq 0$, \refE{pv} recovers the Poor-Verd\'u bound~\cite[Th.~1]{PoorVer95}. For $\qy=\qy^{\star}$ in \refE{qyMAP-def}, optimizing over $\gamma\geq 0$, \refE{pv} provides an expression similar to those in Theorem~\ref{thm:mht-alpha}:
\begin{align}
  \bar\epsilon = \max_{\gamma\geq 0} \left\{(1-\gamma) \Pr\left[ \frac{ \pyv(Y|V) \pv(V)}{  \qy^{\star}(Y) } \leq \gamma \right] \right\}.
\label{eqn:tight-pv}
\end{align}

\subsubsection{Lossy source coding}
Finally, we consider a fixed-length lossy compression scenario, for which a converse based on hypothesis testing was recently obtained in~\cite[Th.~8]{Kost12}. The output of a general source $v$ with distribution $\pv$ is mapped to a codeword $w$ in a codebook $\Cc = \{w_1, w_2, \ldots, w_M\}$ with $w_1, w_2, \ldots, w_M$ belonging to the reconstruction alphabet $\Wc$. We define a non-negative real-valued distortion measure $d(v,w)$ and a maximum allowed distortion $D$. The excess distortion probability is thus defined as $\epsilon_d(\Cc, D)\triangleq \Pr\bigl[ d(V,W) > D \bigr]$.
Consider an encoder that maps the source message $v$ to codeword $w$ with smallest pairwise distortion. The distortion associated to the source message $v$ is then
\begin{equation}
  d(v,\Cc) \triangleq \min_{w\in\Cc} d(v,w).
\end{equation}
Consequently, the excess distortion probability is given by
\begin{align} 
  \epsilon_d(\Cc, D) = \sum_{v} \pv(v) \openone\bigl\{ d(v,\Cc) > D \bigr\}.
  \label{eqn:lsc-eps-2}
\end{align} 

Given the possible overlap between covering regions, there is no straightforward equivalence between the excess distortion probability and the error probability of an $M$-ary hypothesis test. We may yet define an alternative binary hypothesis test as follows. Given an observation $v$, we choose $\Hc_0$ if the encoder meets the maximum allowed distortion and $\Hc_1$ otherwise, i.e. the test is defined as
\begin{align}\label{eqn:lsc-test}
  T_{\text{LSC}}(0|v) 
    = \openone\bigl\{ d(v,\Cc) \leq D \bigr\}.
\end{align}
Particularizing \refE{bht-type0error} and \refE{bht-type1error} with this test, yields 
\begin{align}
  \epsilon_{0}(\pv, T_{\text{LSC}})
    &= \sum_{v} \pv(v) \openone\bigl\{ d(v,\Cc) > D \bigr\},
  \label{eqn:lsc-error-0}\\
  \epsilon_{1}(\qv, T_{\text{LSC}})
    &= \sum_{v} \qv(v) \openone\bigl\{ d(v,\Cc) \leq D \bigr\}\\
    &= \QQ[ d(V,\Cc) \leq D ],
  \label{eqn:lsc-error-1}
\end{align}
where $\QQ[\Ec]$ denotes the probability of the event $\Ec$ with respect to the underlying distribution $\qv$.

As \refE{lsc-eps-2} and \refE{lsc-error-0}
coincide, $\epsilon_d(\Cc,D)$ can be lower-bounded by the type-$0$ error of a Neyman-Pearson test, i.e.,
\begin{align}\label{eqn:lsc-bound}
  \epsilon_d(\Cc,D) \geq \max_{\qv} \Bigl\{
    \alpha_{\QQ[ d(V,\Cc) \leq D ]} \bigl(\pv, \qv\bigr) \Bigr\}.
\end{align}
Moreover, \refE{lsc-bound} holds with equality, as the next result shows.
\begin{theorem}\label{thm:lossymetaconverseistight} 
The excess distortion probability of lossy source coding with codebook $\Cc$ and maximum distortion $D$ satisfies
\begin{align}
  \epsilon_d(\Cc,D)
    &= \max_{\qv}
      \Bigl\{\alpha_{\QQ[ d(V,\Cc) \leq D ]}
         \bigl(\pv, \qv\bigr) \Bigr\}
    \label{eqn:lsc-bound-tight}\\
    &\geq \max_{\qv}
      \Bigl\{\alpha_{M \sup_{w\in\Wc} \QQ\left[d(V,w) \leq D \right]}
         \bigl(\pv, \qv\bigr) \Bigr\}.
    \label{eqn:lsc-kostinathm8}
\end{align}
\end{theorem} 
\begin{IEEEproof}
See Appendix \ref{apx:prooflossymetaconverseistight}.
\end{IEEEproof}

The right-hand-side of \refE{lsc-bound-tight} still depends on the codebook $\Cc$ through $\QQ[ d(V,\Cc) \leq D ]$. This dependence disappears in the relaxation \refE{lsc-kostinathm8}, recovering the converse bound in \cite[Th.~8]{Kost12}.
The weakness of \refE{lsc-kostinathm8} comes from relaxing the type-$1$ error in the bound to $M$ times the type-$1$-error contribution of the best possible codeword belonging to the reconstruction alphabet.

In almost-lossless coding, $D=0$, the error events for different codewords no longer overlap, and the problem naturally fits into the hypothesis testing paradigm. Moreover, when $\qv$ is assumed uniform we have that $\QQ\left[  d(V,w) \leq 0 \right] = \QQ\left[ V = w \right] = \frac{1}{|\Vc|}$ for any $w$ and, therefore, \refE{lsc-kostinathm8} is an equality.

\section*{Acknowledgement}
The authors would like to thank Sergio Verd\'u for multiple discussions. We would also thank Te Sun Han for providing the classical version of the Hayashi-Nagaoka's lemma.

\appendices 

\section{Proof of Corollary \ref{cor:mht-mm}} 
\label{apx:mht-mm}

For a binary hypothesis testing problem between the distributions $\pvy$ and $\qvy^{(q)}$  in \refE{qvy-def-mm} we define the test $T_q(0|v,y) \triangleq P^{(q)}_{\hat{V}|Y}(v|y)$.
We now show that the test $T_q$
achieves the same type-I and type-II error probability
as a NP test $T_{\text{NP}}$ in \refE{bht-NPtest}.
To this end, let us fix $\gamma = \mu'$ and
\begin{align}
  p & = \frac{\sum_{y} \sum_{v\in\Sc_{q}(y)} \frac{1}{|\Sc_{q}(y)|} \pvy(v,y) }
                   {\sum_{y} \sum_{v\in\Sc_{q}(y)} \pvy(v,y)}\label{eqn:mr-p0map-def1}\\
      & = \frac{\sum_{y} \sum_{v\in\Sc_{q}(y)} \frac{1}{|\Sc_{q}(y)|} \qvy^{(q)}(v,y) }
                   {\sum_{y} \sum_{v\in\Sc_{q}(y)} \qvy^{(q)}(v,y)},\label{eqn:mr-p0map-def2}     
\end{align}
where equality between \eqref{eqn:mr-p0map-def1} and \eqref{eqn:mr-p0map-def2} holds since $\pvy(v,y) = \mu' \qvy^{(q)}(v,y)$ for all $y$, $v \in \Sc_{q}(y)$.

The type-$0$ error probability of the NP test \refE{bht-NPtest} with these values of $\gamma$ and $p$ is given by
\begin{align}
 \epsilon_0(\pvy, T_{\text{NP}})
    &= 1 - \sum_{v,y}  \pvy(v,y) T_{\text{NP}}(0|v,y)
    \label{eqn:mr-typeI-1}\\
    &= 1 - \sum_{y} \sum_{v \in \Sc_{q}(y)} p \pvy(v,y)
    \label{eqn:mr-typeI-2}\\
    &= 1 - \sum_{y} \sum_{v \in \Sc_{q}(y)} \frac{1}{|\Sc_{q}(y)|} \pvy(v,y)
    \label{eqn:mr-typeI-3}\\
    &= 1 - \sum_{v,y} \pvy(v,y) T_q(0|v,y)
    \label{eqn:mr-typeI-4}\\
    &=  \epsilon_0(\pvy, T_{q}),
    \label{eqn:mr-typeI-5}
\end{align}
where in \eqref{eqn:mr-typeI-2} we used the definition of $T_{\text{NP}}$ in \refE{bht-NPtest} with $P\leftarrow \pvy$ and $Q\leftarrow  \qvy^{(q)}$ and the definition of $\Sc_{q}(y)$ in \refE{qvy-def-mm}; \eqref{eqn:mr-typeI-3} follows from \eqref{eqn:mr-p0map-def1}, and \eqref{eqn:mr-typeI-4} follows from the definition of $T_q$. Analogously, the type-$1$ error probability of the NP test is
\begin{align}
 \epsilon_1(\qvy^{(q)}, T_{\text{NP}})
    &= \sum_{y} \sum_{v \in \Sc_{q}(y)} p \qvy^{(q)}(v,y)
    \label{eqn:mr-typeII-1}\\
    &= \sum_{y} \sum_{v \in \Sc_{q}(y)} \frac{1}{|\Sc_{q}(y)|} \qvy^{(q)}(v,y)
    \label{eqn:mr-typeII-2}\\
    &= \sum_{v,y} \qvy^{(q)}(v,y) T_q(0|v,y)
    \label{eqn:mr-typeII-3}\\
    &=   \epsilon_1(\qvy^{(q)}, T_q),
    \label{eqn:mr-typeII-4}
\end{align}
where 
\eqref{eqn:mr-typeII-2} follows from \eqref{eqn:mr-p0map-def2};
and \eqref{eqn:mr-typeII-3} follows from the definition of $T_q$.

Then, using \eqref{eqn:mr-typeI-1}-\eqref{eqn:mr-typeI-5} and
\eqref{eqn:mr-typeII-1}-\eqref{eqn:mr-typeII-4}, we obtain
\begin{align}
    \alpha_{\epsilon_{1}\bigl(\qvy^{(q)},T_{q}\bigr)} \bigl(\pvy,\qvy^{(q)}\bigr) &=\epsilon_{0}(\pvy, T_{\text{NP}}) \label{eqn:mr-chain-1}\\
       &=\epsilon_{0}(\pvy, T_{q}).        \label{eqn:mr-chain-2}
\end{align}
Noting that $\bar\epsilon\bigl(P^{(q)}_{\hat{V}|Y}\bigr)$ and $\epsilon_{0}(\pvy, T_{q})$ coincide by definition, then \refE{meta-suboptimal} holds with equality for $\qvy =\qvy^{(q)}$.

Applying Lemma~\ref{lem:alpha-relax-2} to \refE{meta-suboptimal} and fixing $\qvy =\qvy^{(q)}$ yields
\begin{align} 
\bar\epsilon\Bigl(P^{(q)}_{\hat{V}|Y}\Bigr) \geq \sup_{\gamma'\geq 0} \Biggl\{ \Pr\left[ \frac{\pvy(V,Y)}{ \qvy^{(q)}(V,Y) } \leq \gamma' \right] - \gamma' \epsilon_{1}\Bigl(\qvy^{(q)},P^{(q)}_{\hat{V}|Y}\Bigr) \Biggr\}. \label{eqn:mr-chain-3}
\end{align}
Choosing $\gamma' = \mu'$ in \refE{mr-chain-3} direct computation shows
that 
\begin{align} 
  \Pr\left[ \frac{\pvy(V,Y)}{ \qvy^{(q)}(V,Y) } \!\leq\! \mu' \right] 
    &= \Pr\left[ {q(V,Y)}\!\leq\!{\max_{v'} q(v',Y)} \right]
    \label{eqn:mr-chain-4a}\\
    &= 1
    \label{eqn:mr-chain-4b}
\end{align} 
and 
\begin{align} 
\mu'\epsilon_{1}\Bigl(\qvy^{(q)},P^{(q)}_{\hat{V}|Y}\Bigr) 
    &= \sum_{v,y} \pvy(v,y) \frac{\max_{v'} q(v',y)}{q(v,y)} P^{(q)}_{\hat{V}|Y}(v|y)
    \label{eqn:mr-chain-5a}\\
    &= \sum_{v,y} \pvy(v,y) P^{(q)}_{\hat{V}|Y}(v|y),
    \label{eqn:mr-chain-5b}
\end{align} 
where in \refE{mr-chain-5b} we have used that $P^{(q)}_{\hat{V}|Y}(v|y) \neq 0$ implies $q(v,y) = \max_{v'} q(v',y)$. Therefore, substituting \refE{mr-chain-4a}-\refE{mr-chain-4b} and \refE{mr-chain-5a}-\refE{mr-chain-5b} in \refE{mr-chain-3}, and using the definition of $\bar\epsilon(P_{\hat{V}|Y})$ in \refE{mht-epsdef-2}, we conclude that \refE{mr-chain-3} holds with equality, and so does \refE{tight-vh-suboptimal} with $\qvy =\qvy^{(q)}$.

\section{Proof of Lemmas \ref{lem:alpha-relax-2} and \ref{lem:alpha-relax-1}}
\label{apx:alpha-relax}

Consider a binary hypothesis test between  distributions $P$ and $Q$ defined over the alphabet $\Yc$. Let us denote by $\PP[\Ec]$ the probability of the event $\Ec$ with respect to the underlying distribution $P$, and $\QQ[\Ec]$ that with respect to $Q$.

For the sake of clarity we assume that, for a given type-$1$ error $\beta$, the term $p$ in \refE{bht-NPtest} is equal to zero. The proof easily extends to arbitrary $p$, although with more complicated notation. Then, there exists $\gamma^{\star}$ such that
\begin{align}\label{eqn:betaNPtest}
 \beta = \QQ\left[  \frac{P(Y)}{Q(Y)} > \gamma^{\star} \right],
\end{align}
and the NP lemma yields 
\begin{align}\label{eqn:alphaNPtest}
 \alpha_{\beta}(P,Q) = \PP\left[  \frac{P(Y)}{Q(Y)} \leq \gamma^{\star} \right].
\end{align}

For $0 \leq \gamma' < \gamma^{\star}$, $\PP\left[  \frac{P(Y)}{Q(Y)} \leq \gamma' \right] \leq \PP\left[  \frac{P(Y)}{Q(Y)} \leq \gamma^{\star} \right] = \alpha_{\beta}(P,Q)$. Then both Lemmas \ref{lem:alpha-relax-2} and \ref{lem:alpha-relax-1} hold trivially. 

For $\gamma' \geq \gamma^{\star}$ it follows that
\begin{align}
\alpha_{\beta}(P,Q)
    &= \PP\left[  \frac{P(Y)}{Q(Y)}\!\leq\! \gamma' \right] 
     - \PP\left[ \gamma^{\star}\!<\!\frac{P(Y)}{Q(Y)} \leq \gamma' \right]\label{eqn:proofbound2_1} \\
   &\geq \PP\left[  \frac{P(Y)}{Q(Y)}\!\leq\! \gamma' \right] 
     - \gamma' \QQ\left[ \gamma^{\star}\!<\! \frac{P(Y)}{Q(Y)} \leq \gamma' \right]
\label{eqn:proofbound2_2}\\
   &= \PP\left[  \frac{P(Y)}{Q(Y)}\!\leq\!\gamma' \right] 
     - \gamma' \Biggl( \QQ\left[ \frac{P(Y)}{Q(Y)} > \gamma^{\star} \right]
     - \QQ\left[\frac{P(Y)}{Q(Y)} > \gamma' \right] \Biggr), \label{eqn:proofbound2_3}
\end{align}
where \refE{proofbound2_2} follows by noting that in the interval considered $P(y) < \gamma' Q(y)$.
Lemma \ref{lem:alpha-relax-2} follows from \refE{proofbound2_3} by lower bounding $\QQ\left[\frac{P(Y)}{Q(Y)} > \gamma' \right] \geq 0$ and using \refE{betaNPtest}.
In order to prove Lemma \ref{lem:alpha-relax-1},
we shall use in \refE{proofbound2_3} the tighter lower bound
\begin{align}
\QQ\left[\frac{P(Y)}{Q(Y)} > \gamma' \right] 
  \geq \beta \PP\left[\frac{P(Y)}{Q(Y)} > \gamma' \right],
\end{align}
which holds by the assumption in \refE{alpha-relax-assump}.

\section{One Test versus Multiple Tests}\label{apx:Marymultiple}

In this appendix, we prove the equivalence between the optimization problems in \refE{meta} and \refE{multi}.
First, note that the argument of the maximization in \refE{multi} can be written in terms of tests $T_v$ for fixed $v$ as
\begin{align}
\sum_v\pv(v)&\alpha_{Q_{\hat{V}}(v)}\bigl(P_{Y|V=v},\qy\bigr)\notag\\
           &= \sum_v \pv(v)
              \min_{T_v: \epsilon_{1}(\qy, T_v) \leq Q_{\hat{V}}(v)}
                   \Big\{ \epsilon_{0}(P_{Y|V=v}, T_v) \Big\}
                   \label{eqn:mht-cor-dem-1}\\
           &= \sum_{v} \pv(v) \max_{\lambda(v) \geq 0} \min_{T_v} \Biggl\{
                 \sum_{y} P_{Y|V}(y|v) T_v(1|y) 
                 - \lambda(v) \Biggl( \sum_{y'} \qy(y') T_v(0|y') - Q_{\hat{V}}(v) \Biggr) \Biggr\},
                  \label{eqn:mht-cor-dem-2}
\end{align}
where \refE{mht-cor-dem-1} follows from the definition of $\alpha_{(\cdot)} (\cdot)$, and
in \refE{mht-cor-dem-2} we used the definitions of the type-$0$ and type-$1$ errors and introduced the constraints into the objective by means of the Lagrange multipliers $\lambda(v)$.

Similarly, from \refE{meta} we have that
\begin{align}
&\max_{\qy} \; \alpha_{\frac{1}{M}}\left(\pvy, \qv \times \qy\right)\notag\\
  &= \max_{\qv\times\qy} \alpha_{\epsilon_{1}(\qv\times\qy, T_{\text{MAP}})}\left(\pvy, \qv\times\qy\right)
\label{eqn:mht-cor-dem-3}\\
  &= \max_{\qy} \max_{\eta\geq 0} \max_{\qv} \min_{T} \Biggl\{\sum_{v,y} \pvy(v,y) T(1|v,y) 
  + \eta \Biggl( \sum_{v',y'} \qv(v')\qy(y')  \left(T(0|v',y') - P_{\hat V|Y}^{\text{MAP}}(v'|y') \right) \Biggr)\Biggr\}
\label{eqn:mht-cor-dem-4}\\
  &= \max_{\qy} \sum_v \pv(v) \max_{\bar\lambda(v) \geq 0} \min_{T} \Biggl\{ \sum_{y} \pyv(y|v) T(1|v,y)
  + \bar\lambda(v) \Biggl( \sum_{y'}\qy(y')  T(0|v,y') - Q_{\hat{V}}(v) \Biggr)\Biggr\},
\label{eqn:mht-cor-dem-5}
\end{align}
where \refE{mht-cor-dem-3} follows as $\qv$ uniform is a maximizer of the RHS of \refE{mht-cor-dem-3}; in \refE{mht-cor-dem-4} used the definition of $\alpha_{(\cdot)}(\cdot)$, and introduced the constraint into the objective by means of the Lagrange multiplier $\eta$; and in \refE{mht-cor-dem-5} we rearranged terms and defined
\begin{align}
    \bar\lambda(v) \triangleq \frac{\eta\qv(v)}{\pv(v)}.
\end{align}

The result follows from \refE{mht-cor-dem-2} and \refE{mht-cor-dem-5} by optimizing \refE{mht-cor-dem-2} over $\qy$ and identifying $T(i|v,y) \equiv T_v(i|y)$, $i=0,1$.

\section{Proof of Theorem \ref{thm:lossymetaconverseistight}} 
\label{apx:prooflossymetaconverseistight}

We define
\begin{equation}
  \qv^{\Cc}(v) \triangleq \frac{1}{\mu''} \openone\bigl\{ d(v,\Cc) > D \bigr\},
\end{equation}
with $\mu''$ a normalization constant.


The NP test \refE{bht-NPtest} with $P \leftarrow \pv$, $Q \leftarrow \qv^{\Cc}$, $\gamma = \mu''$, $p=1$, particularizes to
\begin{align}\label{eqn:lossyNPtest}
  T_{\text{NP}} (0|v) =
  \begin{cases}
    1, & \text{ if }  \pv(v) \geq \openone\bigl\{ d(v,\Cc) > D \bigr\},\\
    0, & \text{ otherwise}.
  \end{cases}
\end{align}
Assuming that $\pv(v)<1$ for all $v$, eq. \refE{lossyNPtest} reduces to
\begin{align}\label{eqn:lossyNPtestbis}
  T_{\text{NP}} (0|v)
    &= \openone\bigl\{ d(v,\Cc) \leq D \bigr\}\\
    &= T_{\text{LSC}}(0|v).
\end{align}
That is, for  $\qv=\qv^{\Cc}$, the test $T_{\text{LSC}}$ defined in \refE{lsc-test} is optimal in the Newman-Pearson sense. Then it holds that
\begin{align}
\max_{\qv} \left\{
    \alpha_{  \epsilon_{1}(\qv, T_{\text{LSC}}) } \bigl(\pv, \qv\bigr) \right\}
    &\geq  \alpha_{  \epsilon_{1}(\qv^{\Cc}, T_{\text{LSC}}) } \bigl(\pv, \qv^{\Cc}\bigr)
    \label{eqn:lossy-chain-1}\\
    &= \epsilon_{0}\bigl(\pv, T_{\text{LSC}}\bigr)
    \label{eqn:lossy-chain-2}\\
    &= \epsilon_d(\Cc,D),
    \label{eqn:lossy-chain-3}
\end{align}
where the last step follows since \refE{lsc-eps-2} and \refE{lsc-error-0}
coincide.

From \refE{lsc-bound} and \refE{lossy-chain-1}-\refE{lossy-chain-2}, the equality
\refE{lsc-bound-tight} follows by noting that
$\epsilon_{1}(\qv, T_{\text{LSC}}) = \QQ[ d(V,\Cc) \leq D ]$.

Let $P_{W|V}$ denote the encoder that maps the source message $v$ to the codeword $w\in\Cc$ with smallest pairwise distortion. The lower bound \refE{lsc-kostinathm8} follows from the fact that
\begin{align}
  \epsilon_{1}(\qv, T_{\text{LSC}})
    &=\sum_{v} \qv(v) \openone\left\{ d(v,\Cc) \leq D \right\}
    \label{eqn:lossy-epsII-1}\\
    &=\sum_{v} \qv(v) \sum_{w} P_{W|V}(w|v)
     \openone\left\{  d(v,w) \leq D \right\}
    \label{eqn:lossy-epsII-2}\\
    &\leq  \sum_{w\in\Cc}  \sum_{v} \qv(v) \openone\left\{  d(v,w) \leq D \right\}
    \label{eqn:lossy-epsII-3}\\
    &\leq M \sup_{w\in\Cc} \sum_{v} \qv(v) \openone\left\{  d(v,w) \leq D \right\}
    \label{eqn:lossy-epsII-4}\\
    &\leq M \sup_{w\in\Wc} \sum_{v} \qv(v) \openone\left\{  d(v,w) \leq D \right\},
    \label{eqn:lossy-epsII-5}    
\end{align}
where in \refE{lossy-epsII-3} we used that $P_{W|V}(w|v)=0$ for $w\notin\Cc$
and that $P_{W|V}(w|v)\leq 1$ for $w\in\Cc$; \refE{lossy-epsII-4} follows from considering the largest term in the sum, and in \refE{lossy-epsII-5} we relaxed the set over which the maximization is performed. 

\balance

\bibliographystyle{IEEEtran}
\bibliography{bib/references}

\end{document}

%% file: figs/bounds-cc.tikz
%
%
\begin{tikzpicture}

\begin{axis}[%
width=4in,
height=3in,
scale only axis,
separate axis lines,
every outer x axis line/.append style={black},
every x tick label/.append style={font=\color{black}},
xmin=0,
xmax=25,
xlabel={Blocklength, $n$},
xmajorgrids,
every outer y axis line/.append style={black},
every y tick label/.append style={font=\color{black}},
ymode=log,
ymin=2e-05,
ymax=1,
yminorticks=true,
ylabel={Error probability, $\bar\epsilon$},
ymajorgrids,
yminorgrids,
legend style={legend cell align=left,align=left,draw=black}
]
\addplot [color=black,solid,line width=1.0pt]
  table[row sep=crcr]{%
2	0.19\\
3	0.19\\
4	0.1252\\
5	0.06688\\
6	0.0552160000000001\\
7	0.0394696\\
8	0.0221485599999998\\
9	0.0178970320000002\\
10	0.0127951983999999\\
11	0.0074382731199999\\
12	0.00590772303999967\\
13	0.00424324982799886\\
14	0.00250416229960129\\
15	0.00197899230340031\\
16	0.00142090017497848\\
17	0.000848881571380608\\
18	0.000667674160902991\\
19	0.000480107475540254\\
20	0.000288889429572492\\
21	0.000226816689471302\\
22	0.000163176904789788\\
23	9.87667928989583e-05\\
24	7.74293995697557e-05\\
25	5.57564937623362e-05\\
26	0\\
27	0\\
28	0\\
29	0\\
30	0\\
31	0\\
32	0\\
33	0\\
34	0\\
35	0\\
36	0\\
37	0\\
38	0\\
39	0\\
40	0\\
41	0\\
42	0\\
43	0\\
44	0\\
45	0\\
46	0\\
47	0\\
48	0\\
49	0\\
50	0\\
51	0\\
52	0\\
53	0\\
54	0\\
55	0\\
56	0\\
57	0\\
58	0\\
59	0\\
60	0\\
61	0\\
62	0\\
63	0\\
64	0\\
65	0\\
66	0\\
67	0\\
68	0\\
69	0\\
70	0\\
71	0\\
72	0\\
73	0\\
74	0\\
75	0\\
76	0\\
77	0\\
78	0\\
79	0\\
80	0\\
81	0\\
82	0\\
83	0\\
84	0\\
85	0\\
86	0\\
87	0\\
88	0\\
89	0\\
90	0\\
91	0\\
92	0\\
93	0\\
94	0\\
95	0\\
96	0\\
97	0\\
98	0\\
99	0\\
100	0\\
101	0\\
102	0\\
103	0\\
104	0\\
105	0\\
106	0\\
107	0\\
108	0\\
109	0\\
110	0\\
111	0\\
112	0\\
113	0\\
114	0\\
115	0\\
116	0\\
117	0\\
118	0\\
119	0\\
120	0\\
121	0\\
122	0\\
123	0\\
124	0\\
125	0\\
126	0\\
127	0\\
128	0\\
129	0\\
130	0\\
131	0\\
132	0\\
133	0\\
134	0\\
135	0\\
136	0\\
137	0\\
138	0\\
139	0\\
140	0\\
141	0\\
142	0\\
143	0\\
144	0\\
145	0\\
146	0\\
147	0\\
148	0\\
149	0\\
150	0\\
};
\addlegendentry{Best code};

\addplot [color=black,line width=1.0pt,only marks,mark=o,mark options={solid}]
  table[row sep=crcr]{%
2	0.190528068544032\\
3	0.19\\
4	0.1252\\
5	0.0668800000000003\\
6	0.0552160000000007\\
7	0.0394695999999999\\
8	0.0221485599999977\\
9	0.0178970319999959\\
10	0.0127951984000239\\
11	0.00743827311999401\\
12	0.00590772303998954\\
13	0.00424324982790503\\
14	0.00250416229975614\\
15	0.00197899230301124\\
16	0.00142090017579388\\
17	0.000848881571390958\\
18	0.000667674162529774\\
19	0.000480107464421287\\
20	0.000288889419489902\\
21	0.000226816686739295\\
22	0.000163176910548235\\
23	9.87668866165858e-05\\
24	7.74293441506702e-05\\
25	5.57565396467341e-05\\
26	0\\
27	0\\
28	0\\
29	0\\
30	0\\
31	0\\
32	0\\
33	0\\
34	0\\
35	0\\
36	0\\
37	0\\
38	0\\
39	0\\
40	0\\
41	0\\
42	0\\
43	0\\
44	0\\
45	0\\
46	0\\
47	0\\
48	0\\
49	0\\
50	0\\
51	0\\
52	0\\
53	0\\
54	0\\
55	0\\
56	0\\
57	0\\
58	0\\
59	0\\
60	0\\
61	0\\
62	0\\
63	0\\
64	0\\
65	0\\
66	0\\
67	0\\
68	0\\
69	0\\
70	0\\
71	0\\
72	0\\
73	0\\
74	0\\
75	0\\
76	0\\
77	0\\
78	0\\
79	0\\
80	0\\
81	0\\
82	0\\
83	0\\
84	0\\
85	0\\
86	0\\
87	0\\
88	0\\
89	0\\
90	0\\
91	0\\
92	0\\
93	0\\
94	0\\
95	0\\
96	0\\
97	0\\
98	0\\
99	0\\
100	0\\
101	0\\
102	0\\
103	0\\
104	0\\
105	0\\
106	0\\
107	0\\
108	0\\
109	0\\
110	0\\
111	0\\
112	0\\
113	0\\
114	0\\
115	0\\
116	0\\
117	0\\
118	0\\
119	0\\
120	0\\
121	0\\
122	0\\
123	0\\
124	0\\
125	0\\
126	0\\
127	0\\
128	0\\
129	0\\
130	0\\
131	0\\
132	0\\
133	0\\
134	0\\
135	0\\
136	0\\
137	0\\
138	0\\
139	0\\
140	0\\
141	0\\
142	0\\
143	0\\
144	0\\
145	0\\
146	0\\
147	0\\
148	0\\
149	0\\
150	0\\
};
\addlegendentry{Theorem 1, \eqref{eqn:cc-bound-1}};

\addplot [color=blue,dash pattern=on 1pt off 3pt on 3pt off 3pt,line width=1.0pt]
  table[row sep=crcr]{%
2	0.190528068544032\\
3	0.19\\
4	0.1252\\
5	0.0668799999999998\\
6	0.0552159999999997\\
7	0.0237231999999998\\
8	0.0221485599999998\\
9	0.00939397599999969\\
10	0.00854367039999981\\
11	0.00514244799999963\\
12	0.00322926039999971\\
13	0.00234919410399959\\
14	0.00120587319423948\\
15	0.00099132833677551\\
16	0.00044677009743721\\
17	0.000400558581509225\\
18	0.000164628801800015\\
19	0.000157565971317108\\
20	6.49592091687134e-05\\
21	6.08786672257056e-05\\
22	3.32326566456764e-05\\
23	2.3225025855389e-05\\
24	1.51115528110868e-05\\
25	8.77775594199104e-06\\
26	0\\
27	0\\
28	0\\
29	0\\
30	0\\
31	0\\
32	0\\
33	0\\
34	0\\
35	0\\
36	0\\
37	0\\
38	0\\
39	0\\
40	0\\
41	0\\
42	0\\
43	0\\
44	0\\
45	0\\
46	0\\
47	0\\
48	0\\
49	0\\
50	0\\
51	0\\
52	0\\
53	0\\
54	0\\
55	0\\
56	0\\
57	0\\
58	0\\
59	0\\
60	0\\
61	0\\
62	0\\
63	0\\
64	0\\
65	0\\
66	0\\
67	0\\
68	0\\
69	0\\
70	0\\
71	0\\
72	0\\
73	0\\
74	0\\
75	0\\
76	0\\
77	0\\
78	0\\
79	0\\
80	0\\
81	0\\
82	0\\
83	0\\
84	0\\
85	0\\
86	0\\
87	0\\
88	0\\
89	0\\
90	0\\
91	0\\
92	0\\
93	0\\
94	0\\
95	0\\
96	0\\
97	0\\
98	0\\
99	0\\
100	0\\
101	0\\
102	0\\
103	0\\
104	0\\
105	0\\
106	0\\
107	0\\
108	0\\
109	0\\
110	0\\
111	0\\
112	0\\
113	0\\
114	0\\
115	0\\
116	0\\
117	0\\
118	0\\
119	0\\
120	0\\
121	0\\
122	0\\
123	0\\
124	0\\
125	0\\
126	0\\
127	0\\
128	0\\
129	0\\
130	0\\
131	0\\
132	0\\
133	0\\
134	0\\
135	0\\
136	0\\
137	0\\
138	0\\
139	0\\
140	0\\
141	0\\
142	0\\
143	0\\
144	0\\
145	0\\
146	0\\
147	0\\
148	0\\
149	0\\
150	0\\
};
\addlegendentry{Converse, \eqref{eqn:cc-bound-2}};

\end{axis}
\end{tikzpicture}%